\documentclass[aps,prl,twocolumn,floatfix,superscriptaddress,balancelastpage,nofootinbib,amsmath,preprintnumbers]{revtex4-1}[11pt]
\pdfoutput=1
\usepackage{graphicx}
\usepackage{amsmath,amssymb,mathrsfs}
\usepackage{bbm}
\usepackage{color}
\usepackage{ulem}
\usepackage{dsfont}
\usepackage{cancel}
\usepackage{dsfont}
\usepackage{epstopdf}
\usepackage{epsfig}
\usepackage{bm}
\usepackage{dcolumn}
\usepackage{hyperref}

\newcommand{\mysection}[1]{{\noindent \bf #1.}}

\newcommand{\ben}{\begin{enumerate}}
\newcommand{\een}{\end{enumerate}}
\newcommand{\bit}{\begin{itemize}}
\newcommand{\eit}{\end{itemize}}

\newcommand{\beqa}{\begin{eqnarray}}
\newcommand{\eeqa}{\end{eqnarray}}
\newcommand{\beq}{\begin{equation}}
\newcommand{\eeq}{\end{equation}}
\newcommand{\bay}{\begin{array}}
\newcommand{\eay}{\end{array}}

\def\ifmath#1{\relax\ifmmode #1\else $#1$\fi}

\def\gsim{\ \rlap{\raise 3pt \hbox{$>$}}{\lower 3pt \hbox{$\sim$}}\ }
\def\lsim{\ \rlap{\raise 3pt \hbox{$<$}}{\lower 3pt \hbox{$\sim$}}\ }

\def\ls#1{\ifmath{_{\lower1.5pt\hbox{$\scriptstyle #1$}}}}
\def\lsup#1{^{\lower 6pt\hbox{$\scriptstyle#1$}}}

\def\bracket#1#2 {\mathinner{\langle{#1}|{#2}\rangle}}

\def\bracket#1#2 {\mathinner{\langle{#1}|{#2}\rangle}}

\def\U1D{{{\rm U}(1)_{\rm D}}}

\def\babar{\mbox{\sl B\hspace{-0.4em} {\small\sl A}\hspace{-0.37em} \sl B\hspace{-0.4em} {\small\sl A\hspace{-0.02em}R}}} 

\begin{document}

\title{Strong Constraints on Sub-GeV Dark Sectors from SLAC Beam Dump E137}

\preprint{EFI-14-17, YITP-SB-14-16}

\author{Brian Batell
}
\affiliation{Enrico Fermi Institute, University of Chicago, Chicago, IL 60637
}

\author{Rouven Essig}
\affiliation{C.N.~Yang Institute for Theoretical Physics, Stony Brook University, Stony Brook, NY 11794
}

\author{Ze'ev Surujon}
\affiliation{C.N.~Yang Institute for Theoretical Physics, Stony Brook University, Stony Brook, NY 11794
}

\begin{abstract}
We present new constraints on sub-GeV dark matter and dark photons from the electron beam-dump experiment E137 conducted at 
SLAC in 1980--1982. 
Dark matter interacting with electrons (e.g., via a dark photon) could have been produced in the electron-target collisions 
and scattered off electrons in the E137 detector, producing the striking, zero-background signature of a high-energy electromagnetic shower that points back to the beam dump.
E137 probes new and significant ranges of parameter space, and constrains the well-motivated possibility that 
dark photons that decay to light dark-sector particles 
can explain the $\sim 3.6 \sigma$ discrepancy between the measured and SM value of 
the muon anomalous magnetic moment.  
It also restricts the parameter space in which the relic density of dark matter in these models is obtained from thermal freeze-out. 
E137 also convincingly demonstrates that (cosmic) backgrounds can be controlled and thus serves as a powerful proof-of-principle 
for future beam-dump searches for sub-GeV 
dark-sector particles scattering off electrons in the detector. 
\end{abstract}

\maketitle

\mysection{INTRODUCTION}
Dark matter (DM) with mass below $\sim 1$~GeV and interacting with Standard Model (SM) particles through a light mediator 
is a viable and natural possibility consistent with all known data (see e.g.~\cite{Boehm:2003hm,Boehm:2003ha,BPR,Essig:2011nj,Essig:2010ye,Lin:2011gj,Feng:2008ya,Chu:2011be,Graham:2012su,Strassler:2006im,Davoudiasl:2013jma,Davoudiasl:2014kua}).
High-intensity fixed-target experiments have impressive sensitivity to such light DM~\cite{BPR}. 
The basic experimental strategy begins with the production of a relativistic DM beam out of electron or proton collisions with a fixed target, 
followed by detection via DM scattering in a detector positioned downstream of the target.
The prospects of proton fixed-target experiments, including several ongoing neutrino oscillation experiments, have been investigated in~\cite{BPR,deNiverville:2011it,deNiverville:2012ij,Morrissey:2014yma,Batell:2014yra}, and the MiniBooNE experiment at FNAL is presently conducting the first dedicated search~\cite{Dharmapalan:2012xp}. 
More recently, the potential of electron beam-dump experiments has been explored~\cite{Izaguirre:2013uxa,Diamond:2013oda,Izaguirre:2014dua} \cite{Morrissey:2014yma}. 
These proposals complement the ongoing efforts to probe sub-GeV DM with low-energy $e^+ e^-$ colliders~\cite{Essig:2013vha} and direct detection experiments via DM-electron scattering~\cite{Essig:2011nj,Essig:2012yx,Graham:2012su}, as well as broader 
efforts to search for low-mass dark sectors that are weakly coupled to the SM~\cite{Essig:2013lka,Jaeckel:2010ni}. 

\mysection{MODELS}
We focus on a motivated class of DM models based on a new `dark' gauge symmetry, $\U1D$~\cite{Pospelov:2007mp,ArkaniHamed:2008qn,Pospelov:2008jd}, although our discussion applies to 
any scenario in which DM interacts with electrons.  
In this framework, the DM $\chi$ is charged under $\U1D$, which is kinetically mixed with the SM hypercharge, U$(1)_Y$, allowing for DM interactions with the SM~\cite{Holdom,Galison:1983pa}. 
If the $\U1D$ is spontaneously broken, its gauge boson (the `dark photon' $A'$) is massive.  
The low energy effective Lagrangian is 
\begin{eqnarray}
\label{eq:Lagrangian}
\!\!{\cal L} & = &  {\cal L_\chi} - \frac{1}{4} {F'}_{\!\!\! \mu\nu} {F'}^{\mu\nu} +\frac{1}{2}m_{A'}^2 {A'}_{\!\! \mu} {A'}^{\mu} - \frac{\epsilon}{2} {F'}_{\!\!\!\mu\nu}F^{\mu\nu}, \nonumber \\
{\cal L}_\chi & = & 
\begin{cases}
i \bar \chi \not \!\! D \chi - m_\chi \bar \chi \chi,  ~~~~~~~ ({\rm Dirac ~ fermion ~ DM})\\
|D_\mu \chi|^2 - m^2_\chi |\chi|^2,~~~~({\rm Complex ~ scalar ~ DM})
\end{cases} 
\end{eqnarray}
where $D_\mu = \partial_\mu - i g_D {A'}_{\!\! \mu}$ and the dominant mixing is with the SM photon (field 
strength $F_{\mu\nu}$). 
There are four new parameters: the DM mass $m_{\chi}$, the $A'$ mass $m_{A'}$, the dark fine structure constant 
$\alpha_D \equiv g_D^2/4\pi$ ($g_D$ is the $\U1D$ gauge coupling), and the kinetic mixing parameter $\epsilon$. 
We take DM to be either a Dirac fermion or complex scalar. Kinetic mixing is allowed by all symmetries in the effective theory. 
If U(1)$_Y$ is embedded in a Grand Unified Theory (GUT), a characteristic strength 
$\epsilon \sim 10^{-3}-10^{-1}$ ($\sim 10^{-5}-10^{-3}$) is expected if the mixing is generated by a one-(two-)loop 
interaction~\cite{Holdom,Essig:2010ye}. 
In the mass basis (obtained after an appropriate field redefinition in (\ref{eq:Lagrangian})), a small 
coupling of the $A'$ to the electromagnetic current, ${\cal L} \supset -\epsilon e  {A'}_\mu e J^\mu_{EM}$, is induced.

\begin{figure}[t!]
   \begin{center}
  \includegraphics[width=0.49\textwidth]{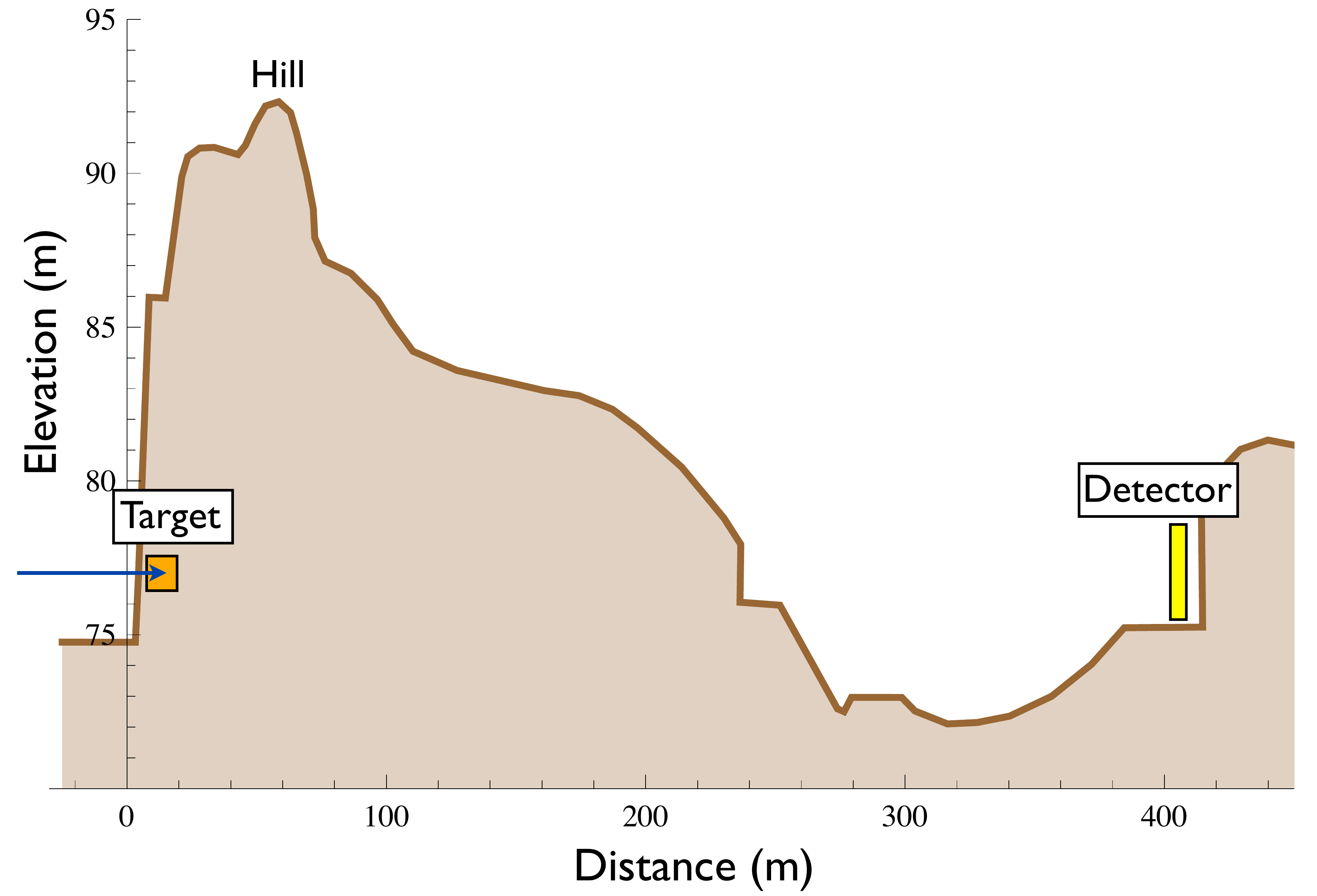} \\
  \vskip 2mm
  \includegraphics[width=0.49\textwidth]{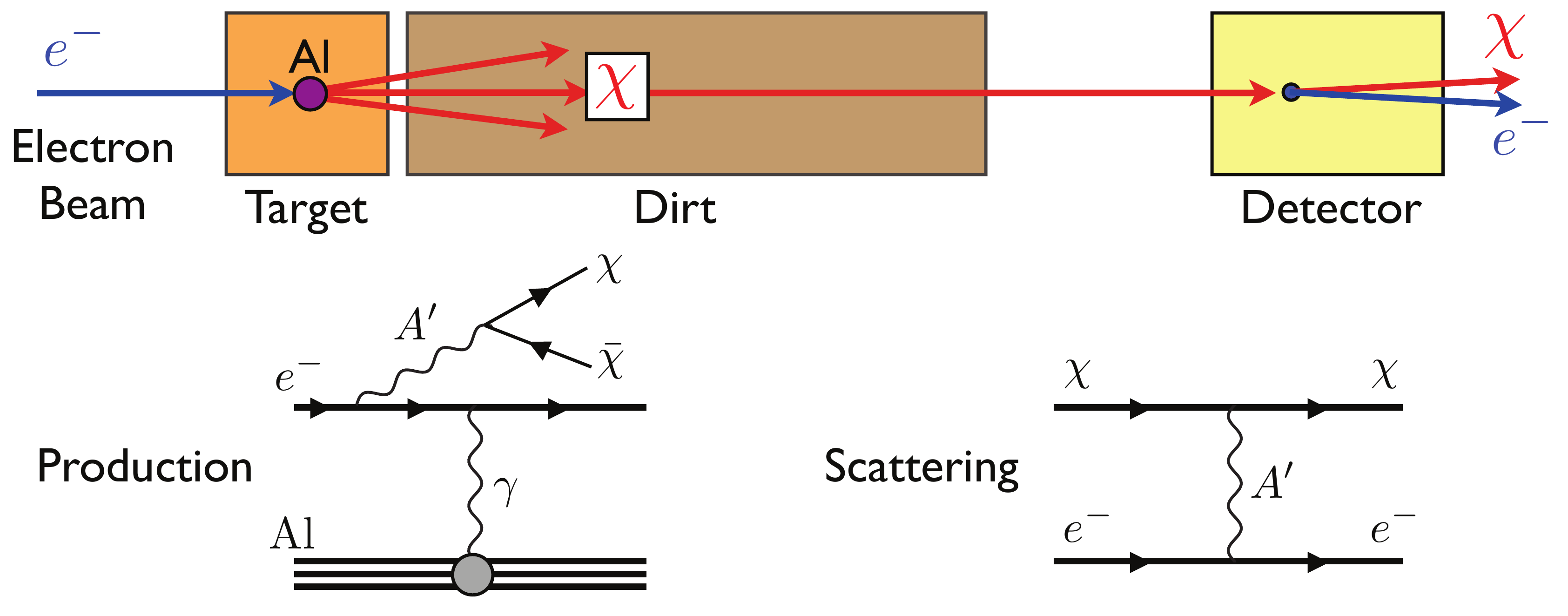}
      \caption{\small 
      {\it Top}: Layout of the E137 experiment (adapted from Fig.~2 in~\cite{E137}). 
      {\it Middle and Bottom}: An electron beam hits an aluminum target, creating DM 
      particles $\chi$ via bremsstrahlung of $A'$ {\it (bottom left)}.  The $\chi$ traverse a $\sim 179$~m deep hill and another 
      $\sim 204$~m-long open region before scattering off electrons {\it (bottom right)}, which are detected in an electromagnetic shower 
      calorimeter.  
      \vspace{-4mm}
       \label{fig:E137}
       }
   \end{center}
\end{figure}
 
We will consider $m_{A'}$ in the MeV to 10~GeV mass range and $m_\chi \lesssim 50$~MeV.  
Several new-physics scenarios can generate naturally a mass for the $A'$ in this range~\cite{ArkaniHamed:2008qp,Cheung:2009qd,Baumgart:2009tn,Morrissey:2009ur,Essig:2009nc}.  
Moreover, over much of this mass range the $A'$ provides a one-loop contribution to the muon anomalous magnetic moment,  
$a_\mu\equiv (g-2)_\mu$, that can account for the $\sim 3.6 \sigma$ discrepancy between its 
measured and SM value~\cite{Bennett:2006fi,Davier:2010nc,Pospelov:2008zw}.  
Various terrestrial, astrophysical, and cosmological tests constrain the scenario~(\ref{eq:Lagrangian})~\cite{Essig:2013lka}.  
We will describe the relevant ones below.  

We emphasize that while~(\ref{eq:Lagrangian}) is an excellent benchmark scenario for sub-GeV DM coupled to a light mediator,
one can easily envision simple extensions or modifications (e.g.~leptophilic DM) to which our discussion is also applicable~\cite{in-progress}.  
We will comment on these in the results section below and discuss how they affect the E137 and various other constraints.

\mysection{SLAC EXPERIMENT E137}
The SLAC experiment E137~\cite{E137} searched for neutral metastable particles produced 
when a 20~GeV electron beam impacted a set of aluminum plates interlaced with cooling water. 
The particles produced at the beam dump needed to traverse 179~m of shielding (provided by a hill) before reaching a 
204~m long open decay region followed by a detector (see Fig.~\ref{fig:E137}, {\it top}).  
The E137 detector consists of an 8-radiation length electromagnetic shower calorimeter 
that can detect charged particles 
or photons produced by the hypothetical particles coming from the dump. 
The detector also employed multiwire proportional chambers to achieve superb angular resolution, rendering it 
sensitive to directional information that was crucial in eliminating (cosmic) background. 
Two experimental runs were performed.  
The lateral dimensions of the detector were 2m $\times$ 3m during Run~1 and
3m $\times$ 3m in Run~2.  The number of electrons on target was $\sim 10$~C ($\sim 20$ C) in Run~1 (Run~2).  

The original analysis in~\cite{E137} searched for axion-like particles decaying to $e^+e^-$, or photinos decaying to a photon and 
gravitino.  
No events were observed that passed quality cuts, pointed back to the dump, and had a shower energy above 1 GeV,
placing strong limits on axions/photinos. 
In \cite{Bjorken:2009mm}, the results were used to set strong constraints on the {\it visible} decay $A' \to e^+e^-$.  

Here, we will use the E137 results to set strong constraints on sub-GeV DM, $\chi$, see Fig.~\ref{fig:E137} ({\it middle} and {\it bottom}).  
We focus on scenarios where $\chi$'s are produced from an on-shell $A'$ that decays {\it invisibly} to $\chi\bar\chi$ or 
via an off-shell $A'$.  
Such $\chi$ inherit a significant portion of the beam energy and travel in the extreme-forward direction; 
an $\mathcal{O}(1)$ fraction of the produced $\chi$ thus intersect the E137 detector and can scatter with electrons in the calorimeter material.  
The ejected electrons will initiate an energetic electromagnetic shower of the type constrained by the E137 search. 
With no observed events, and conservatively assuming no expected background events, we employ a Poisson 95\%~C.L.~limit
of $N_{95}=3$ events. Below, we shall calculate the number of signal events for a fixed $m_\chi$ as function of 
$m_{A'}$, $\epsilon$, and $\alpha_D$, and derive bounds in this parameter space by requiring less than 3 events.

\mysection{SIGNAL RATE CALCULATION}
\begin{figure*}
\begin{center}
\includegraphics[width=0.48\textwidth]{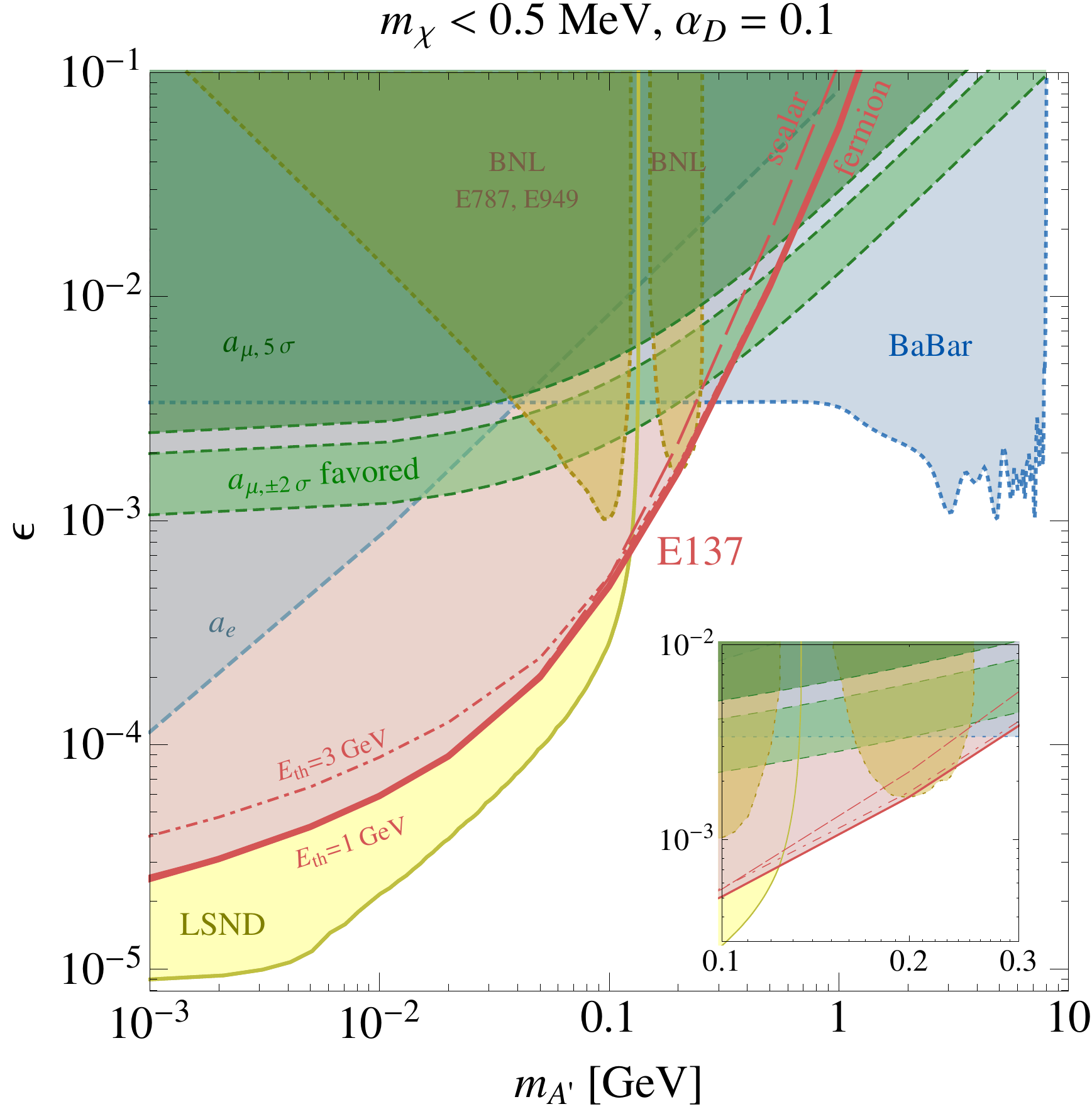}\;
\includegraphics[width=0.48\textwidth]{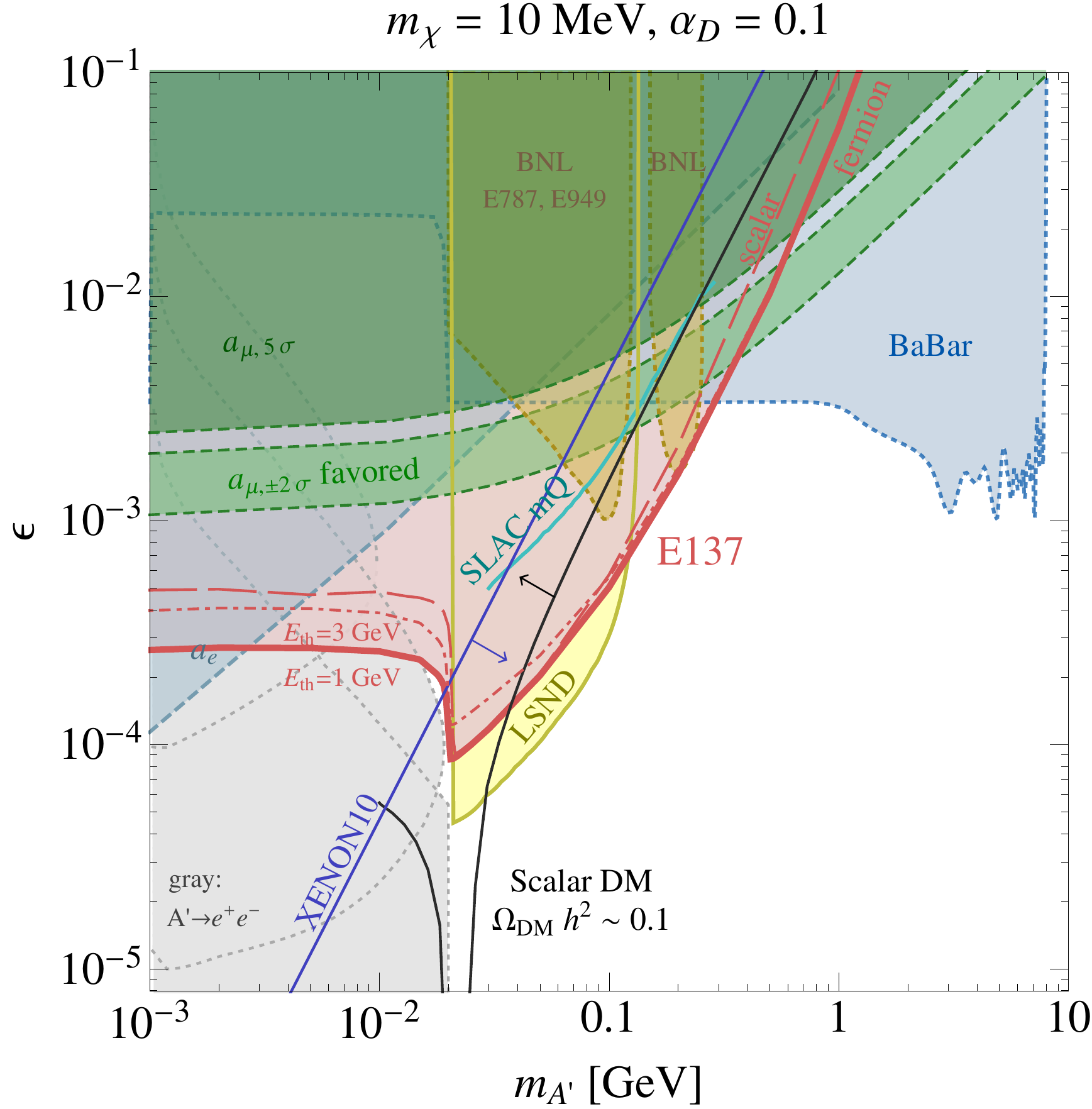}\\
\vskip 5mm
\includegraphics[width=0.48\textwidth]{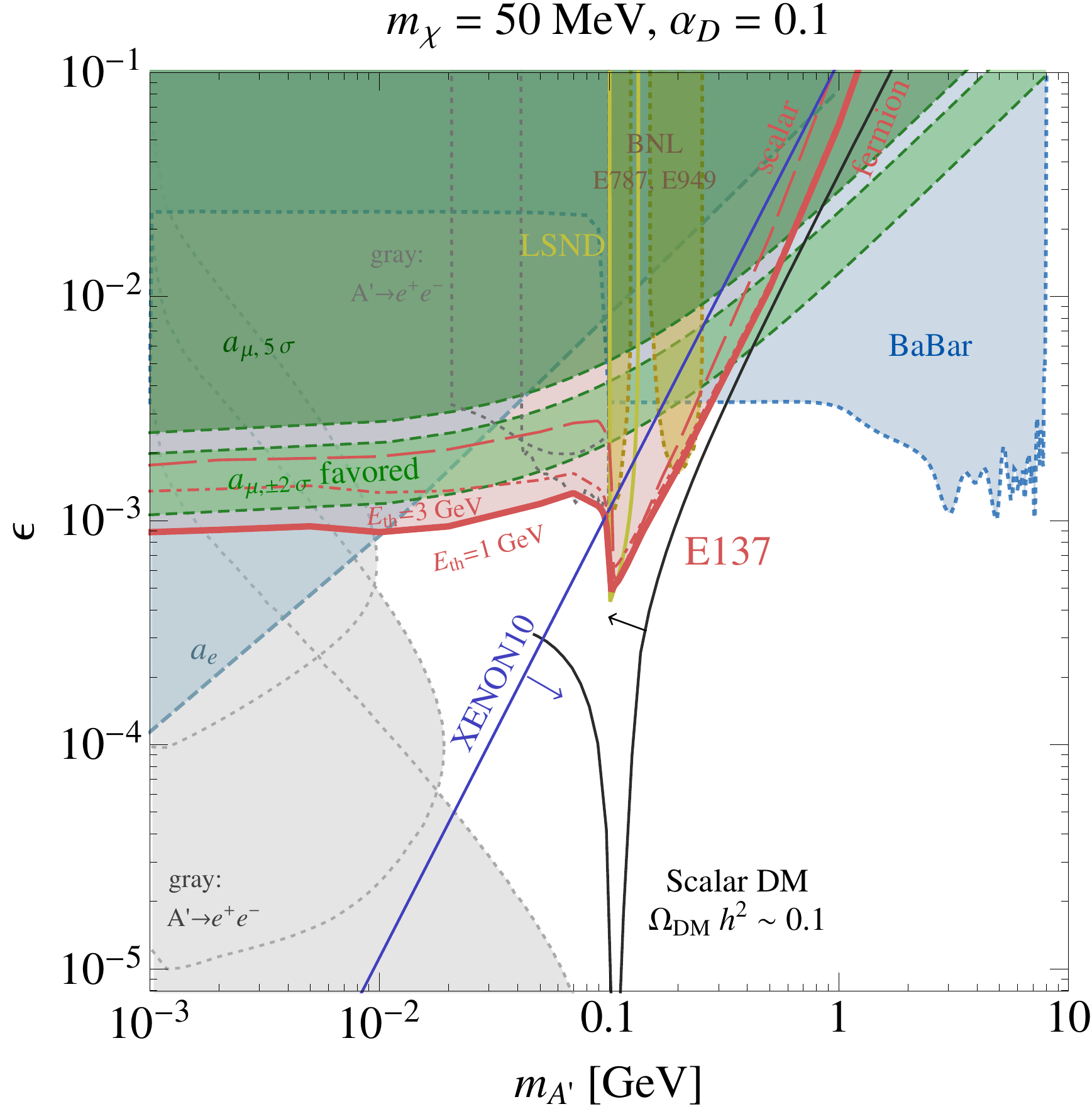}\;
\includegraphics[width=0.472\textwidth]{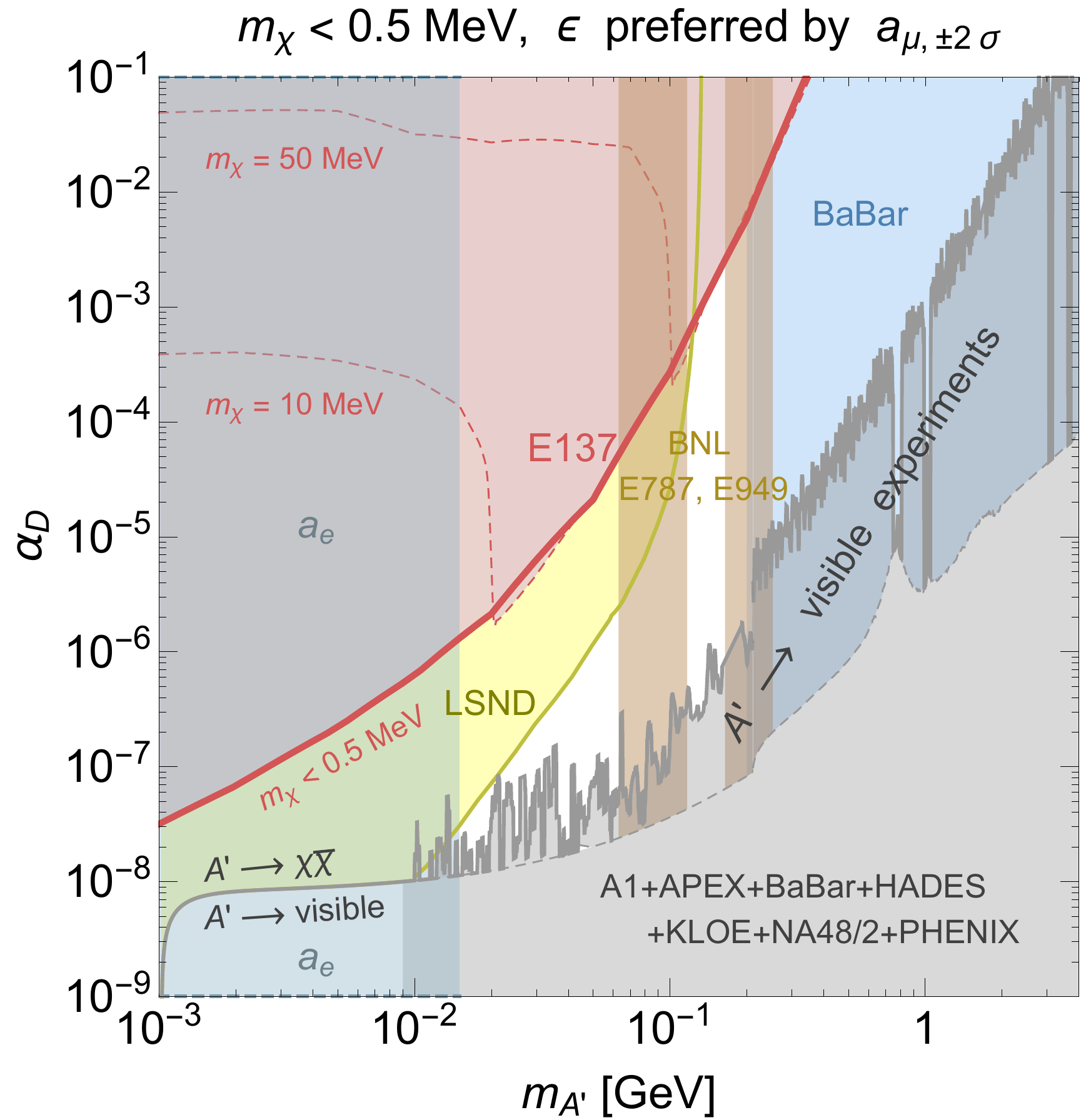}
\end{center}
 \caption{\small 
{\bf Top left:} 
Constraints (95\% C.L.) in the $\epsilon-m_{A'}$ plane for dark photons $A'$ decaying invisibly to light DM $\chi$,  
with $m_\chi < 0.5$~MeV.  
The SLAC E137 experiment excludes a Dirac fermion (red shading/red solid line) 
or complex scalar (red long dashed) DM. 
We fix $\alpha_D = 0.1$ and assume an electron recoil threshold energy of $E_{\rm th} = 1$~GeV 
in the E137 detector (for comparison, the red dotted line shows $E_{\rm th} = 3$~GeV for a fermionic $\chi$).  
Also shown are constraints from the anomalous magnetic moment of the electron 
($a_e$, $2\sigma$, blue dashed) and muon ($a_\mu$, $5\sigma$, dark green dashed), and a 
light-green dashed region in which the $A'$ explains the $a_\mu$ discrepancy. 
Other {\it model-dependent} constraints (see text for details), 
arise from LSND (yellow solid), SLAC mQ experiment (cyan solid), \babar\ (blue dotted), and BNL E787 and E949 (brown dotted). 
The inset focuses on $m_{A'} = 100-300$~MeV.  
{\bf Top right} and {\bf Bottom left:} 
Same as top left but for $m_\chi = 10$~MeV and 50~MeV, respectively. 
Above the black solid line, the thermal relic abundance 
of a scalar $\chi$ satisfies $\Omega_\chi \le \Omega_{\rm DM}$; the region above the blue solid line is excluded if  
$\chi$ can scatter off electrons in the XENON10 experiment, assuming $\chi$ makes up all the DM; the light gray regions/dotted lines 
are excluded from searches for $A' \to e^+e^-$ (if this mode is available for $m_{A'} < 2 m_\chi$) 
in E141, E774, Orsay, HADES, or A1. 
{\bf Bottom right:} 
95\% C.L.~upper limits on $\alpha_D$ as a function of $m_{A'}$ for a Dirac fermion $\chi$, 
assuming $\epsilon$ is fixed to the {\it smallest} value consistent with explaining the $a_\mu$ anomaly. 
The E137 constraint is shown for $m_\chi < 0.5$ MeV (red shading/solid line) and for $m_\chi = 10,50$ MeV (dashed red), while the remaining constraints are only shown for $m_\chi < 0.5$ MeV.  
The solid gray curve is the limit from $A'\to visible$ searches, while the gray dashed represents the transition 
between $A'\to\chi\bar\chi$ and $A'\to visible$ decays dominating. 
}
\label{fig:result}
\end{figure*}
We have employed a Monte-Carlo simulation using \texttt{MadGraph5\_aMC@NLO v2.1.1}~\cite{Alwall:2014hca} to generate DM events produced in electron-aluminum 
nucleus collisions, $e^-N\to e^-N{A'}^{(*)}\rightarrow e^-N \chi\bar\chi$ (where $N$ is a nucleus with $Z=13$, $A=27$), 
and to calculate the total DM production cross section, $\sigma_{\chi \bar \chi}$ 
(we checked all our numerical results against analytic formulas~\cite{Kim:1973he,Bjorken:2009mm,Izaguirre:2013uxa}). 
We include the form factor of the aluminum nucleus~\cite{Bjorken:2009mm,Kim:1973he}, 
which accounts for coherent scattering, as well as nuclear and atomic screening. 
The model~(\ref{eq:Lagrangian}) is implemented using \texttt{FeynRules 2.0}~\cite{Alloul:2013bka}.
We take the thickness of the target to be one radiation length, a reasonable approximation that accounts for 
beam degradation~\cite{Bjorken:2009mm,Izaguirre:2013uxa}.  
The total number of $\chi$ produced is then
\beq
N_\chi = 2 \, \sigma_{\chi\bar \chi} \, N_e \, X_{\rm Al} \, N_A/A_{\rm Al}\,,
\eeq
where $N_e = 30$ C, $X_{\rm Al} = 24.3$ g cm$^{-2}$, $N_A$ is Avogadro's number, 
and $A_{\rm Al} = 26.98$~g$/$mol.  

The fraction of $\chi$ that intersect the detector, $\epsilon_{\rm acc}$, is 
obtained from the Monte-Carlo simulation (and cross-checked analytically) 
by selecting $\chi$ that are produced with angles 
$\tan\theta_{x}< \Delta x/L$ and $\tan\theta_{y}<\Delta y/L$ transverse to the beam direction, 
where $L=383$ m, $\Delta x=1.5$~m, and $\Delta y=1~{\rm m}~(1.5~{\rm m})$  for Run 1 (2).
The angular distribution of scalars $\chi$ produced through an $A'$ is suppressed along the forward direction, 
which results in a lower $\epsilon_{\rm acc}$ compared to fermionic $\chi$~\cite{deNiverville:2012ij,Izaguirre:2013uxa}.
We then take the energy distribution of the DM particles crossing the detector, $(1/N^{\rm acc}_\chi) (dN^{\rm acc}_\chi /dE_\chi)$, and 
convolute it with the $\chi-e^-$ differential scattering cross section, 
\begin{equation}
\frac{d \sigma_{f,s}}{d E_e} = 4 \pi  \epsilon^2 \,  \alpha  \,  \alpha_D 
  \frac{  2 m_e  E_\chi^2 \!  - f_{f,s}(E_e)(E_e - m_e)    }{ (E_\chi^2 - m_\chi^2) ( m_{A'}^2   +2 m_e E_e - 2 m_e^2  )^2 } \,,
\end{equation}
where the subscripts $f,s$ stand for fermion and scalar $\chi$, respectively,
$f_f(E_e) =2 m_e E_\chi - m_e E_e + m_\chi^2 + 2 m_e^2$, 
$f_s =2 m_e E_\chi + m_\chi^2 $, and $E_e$ is the recoil electron energy. 
To conform to the E137 signal region, we impose $E_{e}>E_{\rm th} = 1$ GeV and $\theta_e > 30$ mrad, 
where $\theta_e$ is the angle of the scattered electron, to obtain $\sigma_{\chi e}^{\rm cut}$. 
The number of expected signal events is then given by 
\beq
N_{\chi e} = N_\chi \, \epsilon_{\rm acc} \, \sigma_{\chi e}^{\rm cut} \, \sum_i\,n_{{\rm det}, i} \, L_{{\rm det,i}}\,, 
\eeq
where $n_{{\rm det}, i}$ ($L_{{\rm det},i}$) denotes the $e^-$ number density (length) of detector sub-layer $i$. 
To pass the trigger, $\chi$ must scatter in the first five layers.  
Each layer consists of two sub-layers: 1~cm of plastic scintillator ($n_{\rm det} =  4\times 10^{23}$~cm$^{-3}$), and 
8.9~cm of Al ($n_{\rm det} = 7.8\times 10^{23}$~cm$^{-3}$) in Run~1, or 1.8~cm of Fe 
($n_{\rm det} =  2.2\times 10^{24}$~cm$^{-3}$) in Run 2.   We sum over both runs, weighting by the appropriate fraction of electrons 
dumped in each run. 

Finally, we verified that in the parameter region shown in Fig.~\ref{fig:result} the loss of $\chi$ particles due to scattering in the 
hill is negligible. 

\mysection{RESULTS AND DISCUSSION}
Fig.~\ref{fig:result} ({\it top left})  shows the constraints in the $\epsilon$ versus $m_{A'}$ plane from E137 (red region/lines) 
on the $\U1D$ sub-GeV DM model (\ref{eq:Lagrangian}), assuming $\chi$ is produced in the 
decay of an on-shell $A'$.  The {\it top right} and {\it bottom left} also show $m_\chi = 10$~MeV and $50$~MeV, respectively, 
where the $\chi$ are produced through off-(on-)shell $A'$ for $m_{A'}<2 m_\chi$ ($m_{A'}\ge 2 m_\chi$).
We set $\alpha_D = 0.1$, and note that the limit on $\epsilon$ scales as $\alpha_D^{-1/4}$ ($\alpha_D^{-1/2}$) 
in the on-(off-)shell case.  
We show fermionic and scalar $\chi$ with $E_{\rm th} = 1$~GeV (red solid and long-dashed, respectively), as well as  
fermionic $\chi$ with $E_{\rm th} = 3$~GeV (red dash-dotted); this demonstrates that the result is only mildly sensitive to $E_{\rm th}$.  

It is clear that E137 disfavors a significant and large region in parameter space, probing well into the $\epsilon$ range favored by two-loop GUT mixing up to $m_{A'}\sim 100$~MeV.  
In addition, the $A'$ region favored by the $a_\mu$ anomaly~\cite{Pospelov:2008zw} is significantly constrained from E137 alone.  
Of course, the $a_\mu$ region remains open for small enough $\alpha_D$.  
In the {\it bottom right} of Fig.~\ref{fig:result} we thus show the upper limit on $\alpha_D$ versus $m_{A'}$, 
assuming $\epsilon$ is fixed to be the {\it smallest} value that can explain the 
$a_\mu$ anomaly at $2\sigma$. 
The constraints on $\alpha_D$ are significant, especially for small $m_\chi$. 

The E137 constraints can be evaded in more general sub-GeV DM models.  For example, the constraints disappear in models 
where there is no, or only a reduced, coupling to leptons even if the DM couples to quarks, or where the $\chi$ are not DM but 
decay to lighter particles that are DM and have a negligible coupling to the dark photon.  The constraints are also significantly 
weakened in models in which the connection of the DM to the SM occurs via a higher-dimensional operator.

Several other important constraints exist in the $\epsilon-m_{A'}$ plane.  
While all of these constraints apply to the {\it simplest} $\U1D$ model defined by~(\ref{eq:Lagrangian}), 
most of them can be evaded in more general models, as we will now describe.   
An important limit comes from $a_e$ (blue dashed, same as in~\cite{Essig:2013vha}, see also~\cite{Pospelov:2008zw,Endo:2012hp,Davoudiasl:2012ig}) and $a_\mu$ ($5\sigma$, dark green dashed).  
A more model-dependent limit on $A'\to \chi \bar\chi$ arises from a \babar\ search for 
$e^+e^-\to\gamma + {\rm invisible}$~\cite{Aubert:2008as}, which currently sets the strongest constraint 
above a few hundred MeV (light blue)~\cite{Essig:2013vha,Izaguirre:2013uxa}; improvements from 
Belle 2 (not shown) can be expected if a mono-photon trigger is implemented~\cite{Essig:2013vha}. 
A search by BNL's E787 and E949 for $K\to \pi + {\it invisible}$~\cite{Adler:2004hp,Artamonov:2009sz} 
also constrains $A' \to \chi\bar\chi$ in $K\to \pi + A'$ (brown)~\cite{Dharmapalan:2012xp}; improvements 
would be possible with future searches for rare Kaon decays (not shown)~\cite{Essig:2013vha,ORKA}. 
Prospects from fixed-target experiments like DarkLight~\cite{Freytsis:2009bh,Balewski:2013oza} and VEPP-3~\cite{Wojtsekhowski:2012zq} that are sensitive to invisibly-decaying $A'$ are not shown.  
While the \babar\,, BNL, DarkLight, and VEPP-3 limits/prospects are independent of $\alpha_D$, they can be evaded if the Dirac 
fermion or complex scalar $\chi$ is split into two states $\chi_1$ and $\chi_2$ with 
$m_{\chi_2} > m_{\chi_1} + 2 m_e$; in this case the decay $\chi_2 \to \chi_1 A'^{(*)} \to \chi_1 e^+e^-$ 
could be prompt and change the observed signal~\cite{Izaguirre:2014dua}.  
A careful investigation of this signal in E137 is beyond the scope of this letter~\cite{in-progress}, 
but for small-enough mass splittings it will remain largely unchanged: the resulting beam of $\chi_1$ 
produced at the dump will recoil against an electron and 
up-scatter to $\chi_2$ in the E137 detector.  Even for larger splittings, the $\chi_1$ could scatter off detector nuclei 
into $\chi_2$~\cite{Izaguirre:2014dua}, with the resulting decay $\chi_2\to \chi_1 e^+e^-$ visible in E137.  
However, a search for $\gamma e^+ e^- + {\rm invisible}$ by \babar\ or $\pi e^+ e^- + {\rm invisible}$ by the BNL 
experiments would be able to set constraints on this scenario.  
The BNL constraints can also be evaded if the mediator (not an $A'$) does not couple to quarks.  

The SLAC Millicharge (``mQ'') experiment~\cite{SLACmQ} (cyan solid~\cite{Diamond:2013oda}) sets a 
constraint if $\chi$ interacts with both electrons and quarks. 
The $\chi$ production is similar to E137, but the detection occurs via $\chi$ scattering coherently off {\it nuclei}.
The number of detected $\chi$ depends sensitively on the detector threshold, but could be 
a factor of few larger than in E137; since the existing analysis~\cite{SLACmQ} has not been optimized 
to reduce the relevant backgrounds, the resulting limit is weaker than from E137, which is background free.  
A reanalysis of the SLAC mQ data may improve the constraint~\cite{Diamond:2013oda}, but further study is required to 
determine if it can surpass E137. 

The Liquid Scintillator Neutrino Detector (LSND) sets a strong constraint (yellow region/solid line) if $\chi$ interacts 
with both electrons and quarks (as it would for an $A'$, but not for e.g.~a leptophilic mediator). 
Here $\chi$ is produced through the cascade decays of neutral pions produced in the proton-target collisions, ~$\pi^0\to\gamma A'$, $A' \to \chi\bar\chi$, and detected via its scattering with electrons. 
The resulting constraints are stronger than those from E137 for $m_{A'} < m_{\pi^0}$, assuming the $A'$ decay 
is on-shell~\cite{Auerbach:2001wg,deNiverville:2011it}.  
For $m_\chi > m_{\pi^0}/2$ (not shown), the LSND limits from $\pi^0$ decays disappear, while those from E137 still remain.  

In Fig.~\ref{fig:result} {\it top right} and {\it bottom left}, a scalar $\chi$ satisfies $\Omega_\chi \le \Omega_{\rm DM}$ above the black line.  
The annihilation cross section for $\chi\bar\chi \to A'^* \to e^+e^-$ is $p$-wave 
suppressed at late times~\cite{Boehm:2003hm,deNiverville:2011it,Lin:2011gj}, allowing it to evade strong constraints from the 
Cosmic Microwave Background~\cite{Madhavacheril:2013cna} and gamma-ray searches~\cite{Essig:2013goa}.  
This is perhaps the simplest model that can account for the cosmic DM abundance, and E137 
constrains previously allowed parameter regions of this motivated scenario. 
We also show a blue solid line above which $\chi$ would be disfavored from an 
analysis~\cite{Essig:2012yx} done with a published XENON10 result~\cite{Angle:2011th}.  
Such $\chi$ could scatter off atomic electrons, leading to single- or few-electron events that XENON10 could have detected.  
(This constraint can be evaded if e.g.~$\chi$ does not constitute all the DM or if it is split into two states with a large enough mass splitting.)
Finally, we show in light gray regions/gray dotted lines those areas that have been excluded for searches for 
the {\it visible} decays $A' \to e^+e^-$~\cite{Bjorken:2009mm,Abrahamyan:2011gv,Merkel:2011ze,Andreas:2012mt,Agakishiev:2013fwl,Essig:2010xa,Merkel:2014avp,Babusci:2012cr,Babusci:2014sta,Lees:2014xha,Batley:2015lha}, which is an available mode for $m_{A'} < 2 m_\chi$ in the simplest $\U1D$ model. 
This mode also competes with $A' \to \chi\bar\chi$ for $\alpha_D$ below the solid gray line in the {\it bottom right} 
of Fig.~\ref{fig:result}.  

In summary, E137 is a powerful proof-of-principle that additional beam dump searches for 
DM recoiling off electrons at Jefferson Lab or elsewhere can be successful in mitigating backgrounds.   
It sets significant and unique constraints on sub-GeV DM coupled to a light mediator.  

\vskip -0.2mm
\begin{center} 

{\bf Note added}
\end{center}

The first version of Fig.~\ref{fig:result} ({\it bottom right}) did not include the experimental limits from searches for 
$A'\to visible$ decays above the dashed gray line, where the branching ratio $A'\to\chi\bar\chi$ dominates over the visible mode. 
In this region, the limit on $\alpha_D$ is given by 
\begin{eqnarray}
\alpha_{D, {\rm lim}} = \alpha_{D, {\rm 50\%}} \, \left(\frac{\epsilon^2_{a_\mu}}{\epsilon^2_{\rm lim}} - 1\right)\,, 
\end{eqnarray}
where $\epsilon_{a_\mu}$ is the smallest value that can explain the $a_\mu$ anomaly at $2\sigma$ and $\epsilon_{\rm lim}$ is the 
experimental limit obtained by a search for $A'\to visible$. Also, $N_{\rm eff}$ is the number of visible decay modes available to the $A'$, 
which (in the mass range of interest) is given 
\begin{equation}
N_{\rm eff}(m_{A'}) \simeq h_e(m_{A'})+ h_\mu(m_{A'}) (1+R_{\rm had}(m_{A'})), 
\end{equation}
where $R_{\rm had} = \sigma(e^+e^- \to {\rm hadrons})/\sigma(e^+e^- \to \mu^+\mu^-)$ and 
\begin{equation}
h_f(m_{A'}) = \left(1+\frac{2 m_f^2}{m_{A'}^2}\right)  \sqrt{1-\frac{4 m_f^2}{m_{A'}^2}} \,.
\end{equation}
Finally, the partial widths for $A'\to\chi\bar\chi$ and $A'\to visible$ are equal when $\alpha_D$ is given by
\begin{equation}
\alpha_{D, {\rm 50\%}} = \alpha \ \epsilon_{a_\mu}^2 \ \frac{N_{\rm eff}}{h_\chi}\,,
\end{equation}
where we assumed $\chi$ is a fermion.

\vskip -0.2mm
\begin{center} 

{\bf Acknowledgements}
\end{center}
\vskip -1mm
We are very grateful to J.~D.~Bjorken for useful discussions and correspondence about E$\alpha^{-1}$, 
to J.~A.~Jaros for inspiring questions and discussions, and to O.~Mattelaer for assistance with MadGraph~5.  
We also thank E.~Izaguirre, G.~Krnjaic, W.~Louis, P.~Schuster, and N.~Toro for useful discussions or correspondence.  
We thank Philip Schuster and Natalia Toro for pointing out that the experimental limits from searches for 
$A'\to visible$ decays can extend into the region where the branching ratio $A'\to\chi\bar\chi$ dominates over the $A'\to visible$ 
mode (see Fig.~\ref{fig:result} ({\it bottom right})). 
BB is supported by the NSF under Grant PHY-0756966 and the DOE under Grant DE-SC0003930. 
RE is supported in part by the Department of Energy (DoE) Early Career research program DESC0008061 and by 
a Sloan Foundation Research Fellowship.
ZS is supported in part by the National Science Foundation under Grant PHY-0969739.


\vskip -5mm


\begin{thebibliography}{99}  

\bibitem{Boehm:2003hm} 
  C.~Boehm and P.~Fayet,
  Nucl.\ Phys.\ B {\bf 683}, 219 (2004), 
  hep-ph/0305261.
   
\bibitem{Boehm:2003ha}
C.~Boehm, P.~Fayet and J.~Silk,
Phys.\ Rev.\ D {\bf 69} (2004) 101302, hep-ph/0311143.

\bibitem{BPR} 
  B.~Batell, M.~Pospelov and A.~Ritz,
  Phys.\ Rev.\ D {\bf 80}, 095024 (2009), arXiv:0906.5614. 

\bibitem{Essig:2011nj} 
  R.~Essig, J.~Mardon and T.~Volansky,
  Phys.\ Rev.\ D {\bf 85}, 076007 (2012), arXiv:1108.5383.
  
\bibitem{Essig:2010ye}
R.~Essig, J.~Kaplan, P.~Schuster and N.~Toro,
arXiv:1004.0691.

\bibitem{Lin:2011gj}
T.~Lin, H.~-B.~Yu and K.~M.~Zurek,
Phys.\ Rev.\ D {\bf 85} (2012) 063503, arXiv:1111.0293.

\bibitem{Feng:2008ya} 
  J.~L.~Feng and J.~Kumar,
  Phys.\ Rev.\ Lett.\  {\bf 101}, 231301 (2008), arXiv:0803.4196.

\bibitem{Chu:2011be}
X.~Chu, T.~Hambye and M.~H.~G.~Tytgat,
JCAP {\bf 1205} (2012) 034, arXiv:1112.0493.

\bibitem{Graham:2012su}
P.~W.~Graham {\it et al.}, 
Phys.\ Dark Univ.\ {\bf 1} (2012) 32, arXiv:1203.2531.
   
\bibitem{Strassler:2006im}
M.~J.~Strassler and K.~M.~Zurek,
Phys.\ Lett.\ B {\bf 651} (2007) 374, hep-ph/0604261.

\bibitem{Davoudiasl:2013jma}
H.~Davoudiasl and I.~M.~Lewis,
Phys.\ Rev.\ D {\bf 89} (2014) 055026, arXiv:1309.6640.

\bibitem{Davoudiasl:2014kua}
H.~Davoudiasl, H.~-S.~Lee and W.~J.~Marciano,
Phys.\ Rev.\ D {\bf 89} (2014) 095006, arXiv:1402.3620.

\bibitem{deNiverville:2011it} 
  P.~deNiverville, M.~Pospelov and A.~Ritz,
  Phys.\ Rev.\ D {\bf 84}, 075020 (2011), arXiv:1107.4580.

\bibitem{deNiverville:2012ij} 
  P.~deNiverville, D.~McKeen and A.~Ritz,
  Phys.\ Rev.\ D {\bf 86}, 035022 (2012), arXiv:1205.3499. 
  
\bibitem{Morrissey:2014yma} 
  D.~E.~Morrissey and A.~P.~Spray,
  arXiv:1402.4817. 
  
\bibitem{Batell:2014yra} 
  B.~Batell, P.~deNiverville, D.~McKeen, M.~Pospelov and A.~Ritz,
  arXiv:1405.7049. 
  
\bibitem{Dharmapalan:2012xp} 
  R.~Dharmapalan {\it et al.}  [MiniBooNE Collaboration],
  arXiv:1211.2258. 
    
\bibitem{Izaguirre:2013uxa} 
  E.~Izaguirre, G.~Krnjaic, P.~Schuster and N.~Toro,
  Phys.\ Rev.\ D {\bf 88}, 114015 (2013), arXiv:1307.6554.
  
\bibitem{Diamond:2013oda} 
  M.~D.~Diamond and P.~Schuster,
  Phys.\ Rev.\ Lett.\  {\bf 111}, 221803 (2013), arXiv:1307.6861.
  
\bibitem{Izaguirre:2014dua} 
  E.~Izaguirre, G.~Krnjaic, P.~Schuster and N.~Toro,
  arXiv:1403.6826. 
  
  \bibitem{Essig:2013vha} 
  R.~Essig, J.~Mardon, M.~Papucci, T.~Volansky and Y.~-M.~Zhong,
  JHEP {\bf 1311}, 167 (2013), arXiv:1309.5084.
  
\bibitem{Essig:2012yx} 
  R.~Essig, A.~Manalaysay, J.~Mardon, P.~Sorensen and T.~Volansky,
  Phys.\ Rev.\ Lett.\  {\bf 109}, 021301 (2012), arXiv:1206.2644. 
   
\bibitem{Essig:2013lka} 
  R.~Essig, J.~A.~Jaros, W.~Wester {\it et al.},
  arXiv:1311.0029. 

\bibitem{Jaeckel:2010ni}
J.~Jaeckel and A.~Ringwald,
Ann.\ Rev.\ Nucl.\ Part.\ Sci.\ {\bf 60} (2010) 405, 
arXiv:1002.0329.

\bibitem{Pospelov:2007mp} 
  M.~Pospelov, A.~Ritz and M.~B.~Voloshin,
  Phys.\ Lett.\ B {\bf 662}, 53 (2008), arXiv:0711.4866. 

\bibitem{ArkaniHamed:2008qn} 
  N.~Arkani-Hamed, D.~P.~Finkbeiner, T.~R.~Slatyer and N.~Weiner,
  Phys.\ Rev.\ D {\bf 79}, 015014 (2009), 
arXiv:0810.0713. 

\bibitem{Pospelov:2008jd}
M.~Pospelov and A.~Ritz,
Phys.\ Lett.\ B {\bf 671} (2009) 391,
arXiv:0810.1502.

\bibitem{Holdom} 
  B.~Holdom,
  Phys.\ Lett.\ B {\bf 166}, 196 (1986).

\bibitem{Galison:1983pa}
P.~Galison and A.~Manohar,
Phys.\ Lett.\ B {\bf 136} (1984) 279.

\bibitem{ArkaniHamed:2008qp}
N.~Arkani-Hamed and N.~Weiner,
JHEP {\bf 0812} (2008) 104, 
arXiv:0810.0714.

\bibitem{Cheung:2009qd}
C.~Cheung, J.~T.~Ruderman, L.~T.~Wang and I.~Yavin,
Phys.\ Rev.\ D {\bf 80} (2009) 035008, 
arXiv:0902.3246.

\bibitem{Baumgart:2009tn}
M.~Baumgart, C.~Cheung, J.~T.~Ruderman, L.~T.~Wang and I.~Yavin,
JHEP {\bf 0904} (2009) 014, 
arXiv:0901.0283.

\bibitem{Morrissey:2009ur}
D.~E.~Morrissey, D.~Poland and K.~M.~Zurek,
JHEP {\bf 0907} (2009) 050, 
arXiv:0904.2567.

\bibitem{Essig:2009nc}
R.~Essig, P.~Schuster and N.~Toro,
Phys.\ Rev.\ D {\bf 80} (2009) 015003, 
arXiv:0903.3941. 

\bibitem{E137} 
  J.~D.~Bjorken {\it et al.},
  Phys.\ Rev.\ D {\bf 38}, 3375 (1988).

\bibitem{Bennett:2006fi}
G.~W.~Bennett {\it et al.} [Muon G-2 Collaboration],
Phys.\ Rev.\ D {\bf 73} (2006) 072003, 
hep-ex/0602035.

\bibitem{Davier:2010nc}
M.~Davier, A.~Hoecker, B.~Malaescu and Z.~Zhang,
Eur.\ Phys.\ J.\ C {\bf 71} (2011) 1515
[Erratum-ibid.\ C {\bf 72} (2012) 1874], 
arXiv:1010.4180

\bibitem{Pospelov:2008zw} 
  M.~Pospelov,
  Phys.\ Rev.\ D {\bf 80}, 095002 (2009), arXiv:0811.1030.

\bibitem{in-progress}
B.~Batell, J.~H.~Chang, R.~Essig, Z.~Surujon, in progress.

\bibitem{Bjorken:2009mm}
J.~D.~Bjorken, R.~Essig, P.~Schuster and N.~Toro,
Phys.\ Rev.\ D {\bf 80} (2009) 075018, arXiv:0906.0580.

\bibitem{Alwall:2014hca} 
  J.~Alwall {\it et al.},
  arXiv:1405.0301. 

\bibitem{Kim:1973he}
K.~J.~Kim and Y.~-S.~Tsai,
Phys.\ Rev.\ D {\bf 8} (1973) 3109.

\bibitem{Alloul:2013bka} 
  A.~Alloul {\it et al.} 
  arXiv:1310.1921.

\bibitem{Endo:2012hp}
M.~Endo, K.~Hamaguchi and G.~Mishima,
Phys.\ Rev.\ D {\bf 86} (2012) 095029,[arXiv:1209.2558.

\bibitem{Davoudiasl:2012ig}
H.~Davoudiasl, H.~-S.~Lee and W.~J.~Marciano,
Phys.\ Rev.\ D {\bf 86} (2012) 095009, 
arXiv:1208.2973. 

\bibitem{Aubert:2008as} 
  B.~Aubert {\it et al.}  [BaBar Collaboration],
  arXiv:0808.0017.
  
\bibitem{Adler:2004hp} 
  S.~Adler {\it et al.}  [E787 Collaboration],
  Phys.\ Rev.\ D {\bf 70}, 037102 (2004), 
  hep-ex/0403034.
  
\bibitem{Artamonov:2009sz} 
  A.~V.~Artamonov {\it et al.}  [BNL-E949 Collaboration],
  Phys.\ Rev.\ D {\bf 79}, 092004 (2009), 
arXiv:0903.0030.
  
\bibitem{ORKA} 
  E.~T.~Worcester [ORKA Collaboration],
  PoS KAON {\bf 13}, 035 (2013),
arXiv:1305.7245.

\bibitem{Freytsis:2009bh}
M.~Freytsis, G.~Ovanesyan and J.~Thaler,
JHEP {\bf 1001} (2010) 111, 
arXiv:0909.2862. 

\bibitem{Balewski:2013oza}
J.~Balewski {\it et al.},
arXiv:1307.4432.

\bibitem{Wojtsekhowski:2012zq}
B.~Wojtsekhowski, D.~Nikolenko and I.~Rachek,
arXiv:1207.5089. 

\bibitem{SLACmQ} 
  A.~A.~Prinz {\it et al.},
  Phys.\ Rev.\ Lett.\  {\bf 81}, 1175 (1998),
  hep-ex/9804008.

\bibitem{Auerbach:2001wg} 
  L.~B.~Auerbach {\it et al.}  [LSND Collaboration],
  Phys.\ Rev.\ D {\bf 63}, 112001 (2001), hep-ex/0101039.

\bibitem{Madhavacheril:2013cna}
M.~S.~Madhavacheril, N.~Sehgal and T.~R.~Slatyer,
Phys.\ Rev.\ D {\bf 89} (2014) 103508, arXiv:1310.3815.

\bibitem{Essig:2013goa}
R.~Essig, E.~Kuflik, S.~D.~McDermott, T.~Volansky and K.~M.~Zurek,
JHEP {\bf 1311} (2013) 193, 
arXiv:1309.4091. 

\bibitem{Angle:2011th}
J.~Angle {\it et al.} [XENON10 Collaboration],
Phys.\ Rev.\ Lett.\ {\bf 107} (2011) 051301,
arXiv:1104.3088. 

\bibitem{Abrahamyan:2011gv}
S.~Abrahamyan {\it et al.} [APEX Collaboration],
Phys.\ Rev.\ Lett.\ {\bf 107} (2011) 191804
arXiv:1108.2750. 

\bibitem{Merkel:2011ze}
H.~Merkel {\it et al.} [A1 Collaboration],
Phys.\ Rev.\ Lett.\ {\bf 106} (2011) 251802
[arXiv:1101.4091 [nucl-ex]].

    \bibitem{Andreas:2012mt}
S.~Andreas, C.~Niebuhr and A.~Ringwald,
Phys.\ Rev.\ D {\bf 86} (2012) 095019, 
arXiv:1209.6083.

\bibitem{Agakishiev:2013fwl}
G.~Agakishiev {\it et al.} [HADES Collaboration],
Phys.\ Lett.\ B {\bf 731} (2014) 265, 
arXiv:1311.0216. 

\bibitem{Essig:2010xa}
R.~Essig, P.~Schuster, N.~Toro and B.~Wojtsekhowski,
JHEP {\bf 1102} (2011) 009
[arXiv:1001.2557 [hep-ph]].

\bibitem{Merkel:2014avp}
H.~Merkel {\it et al.},
arXiv:1404.5502. 

\bibitem{Babusci:2012cr}
D.~Babusci {\it et al.} [KLOE-2 Collaboration],
Phys.\ Lett.\ B {\bf 720} (2013) 111, arXiv:1210.3927. 

\bibitem{Babusci:2014sta}
D.~Babusci {\it et al.} [KLOE-2 Collaboration],
arXiv:1404.7772. 

\bibitem{Lees:2014xha} 
  J.~P.~Lees {\it et al.} [BaBar Collaboration],
  Phys.\ Rev.\ Lett.\  {\bf 113}, no. 20, 201801 (2014)
  [arXiv:1406.2980 [hep-ex]].
  
\bibitem{Batley:2015lha} 
  J.~R.~Batley {\it et al.} [NA48/2 Collaboration],
  Phys.\ Lett.\ B {\bf 746}, 178 (2015)
  [arXiv:1504.00607 [hep-ex]].

\end{thebibliography}
\end{document}